\documentclass[12pt]{article}
\usepackage[dvips]{color}
\usepackage{amssymb}
\usepackage{epsfig}

\usepackage[dvips]{color}
\usepackage{epsfig}
\usepackage{amsmath}
\usepackage{graphicx}
\def\be {\begin{equation}}
\def\ee {\end{equation}}
\def\ba {\begin{eqnarray}}
\def\ea {\end{eqnarray}}

\def\bi {\begin{itemize}}
\def\ei {\end{itemize}}
\begin{document}
\begin{large}
\title{\bf  {Generalized Uncertainty Principle Removes The Chandrasekhar Limit}}
\end{large}
\author{Reza Rashidi \\
{\small Department of Physics, Shahid Rajaee Teacher Training University,
Tehran,  IRAN.}\\{\small E-mail: reza.rashidi@srttu.edu}} \maketitle
 \begin{abstract}
The effect of a generalized uncertainty principle on the structure of an ideal white dwarf star is investigated. The equation describing the equilibrium configuration of the star is a generalized form of the Lane-Emden equation. It is proved that the star always has a finite size. It is then argued that the maximum mass of such an ideal white dwarf tends to infinity, as opposed to the conventional case where it has a finite value.
\end{abstract}

\section{Introduction}
The aim of a theory of quantum gravity is the incorporation of quantum mechanics into gravitational theory. At large scale, in the presence of gravity, space-time as the set of all physical events is assumed to be a smooth manifold equipped with a metric tensor. On the other hand, at small scale, as small as atomic scale, the gravitational field is much weaker than the other forces and therefore it is reasonable to ignore it. But at very small distances, as small as the Planck scale, the situation seems different.
 According to the Heisenberg uncertainty principle, an small uncertainty about the position of a particle implies a large uncertainty in momentum space. This large uncertainty also makes an uncertainty in the geometry (i.e. in the metric connections) due to the back reaction effect \cite{1,2}. This space-time fluctuation itself eventually implies an additional uncertainty in the position space. Therefore, we require a new theory which reconciles quantum mechanics and general relativity at least at the Planck scale. At present, in the absence of a reliable theory of quantum gravity, it is reasonable to assume that there is an effective bound on the precision of position measurements. In other words, if we attempt to probe a very small distance or try to infinitely localize a particle, the structure of space-time is then distorted in such a way that we practically acquire a minimum observable length.

Besides this general consideration, many candidates for a theory of quantum gravity such as string theory \cite{3,4,5,6}, loop quantum gravity \cite{7}, and non-commutative geometry \cite{8} predict the existence of a minimal scale or maximal resolution. It can be also derived from various gedanken experiments \cite{9,10,11,12,13}, from black hole physics \cite{14,15}, the holographic principle \cite{16}, and further more \cite{1,17}. Consequently, the existent of a fundamental scale seems to be an integral part of any consistent theory of quantum gravity.

The space-time fluctuations implies an extra uncertainty except the conventional quantum uncertainty. It imposes a change in the Heisenberg uncertainty principle which, at the first approximation, may be written as
\begin{equation}
\triangle x\triangle p\geq\hbar(1+\lambda^2\triangle p ^2), \label{1}
\end{equation}
where $\lambda$ is an unknown dimensional constant \cite{18}. This generalized uncertainty principle (GUP) yields the minimum observable length $\triangle x_{min}=\hbar\lambda$, which is believed has a value close to the Planck length. Thus, $\lambda$ probably has a value around the inverse of the Planck momentum. The generalized uncertainty relation (\ref{1}) can be acquired from the commutation relation
\begin{equation}
[x , p]=i\hbar(1+\lambda^2p^2), \label{a1}
\end{equation}
where $x$ and $p$ respectively denote the position and momentum operators. The natural generalization of (\ref{a1}) in three dimensions which preserves the rotational symmetry is \cite{18}
\begin{eqnarray}
&[x_{i} , p_{j}]&=i\hbar\delta_{ij}(1+\lambda^2\bar{p}^2)\nonumber\\
&[p_{i} , p_{j}]&=0\nonumber\\
&[x_{i}, x_{j}]&=2i\hbar\lambda^2(p_{i}x_{j}-p_{j}x_{i}). \label{a2}
\end{eqnarray}
The last relation leads to a noncommutative geometry.

Another possibility can be obtained by modifying the linear relation between the momentum and wave vector of a particle as $p_{\mu}=\hbar f_{\mu}(k_{\nu})$, where $k_{\nu}$ is the wave vector of the particle and  $f$ is a non-linear injective function. Assuming that the associated operators obey the standard commutation relations
\begin{equation}
[x^{\mu} , x^{\nu}]=0 ,\hspace{.5 cm}
[x^{\mu} , k_{\nu}]=i\delta ^{\mu}_{\nu} ,\hspace{.5 cm}
[k_{\mu}, k_{\nu}]=0,\label{a3}
\end{equation}
one finds
\begin{equation}
[x^{\mu} , x^{\nu}]=0 ,\hspace{.5 cm}
[x^{\mu} , p_{\nu}]=i\hbar\frac{\partial f_{\nu}}{\partial k_{\mu}} ,\hspace{.5 cm}
[p_{\mu}, p_{\nu}]=0.\label{a4}
\end{equation}
These commutation relations then yield the uncertainty relation
\begin{equation}
\triangle x^{i}\triangle p_{i}\geq\frac{\hbar}{2}\langle\frac{\partial f_{i}}{\partial k_{i}}\rangle,
\label{a5}
\end{equation}
which in a suitable limit reproduces GUP \cite{17}.

The generalized uncertainty principle (\ref{1}) leads to many consequences which have been widely studied in detail \cite{17}.  These manifestations of GUP appear at very high energy level or at very small scales, e.g., in a system with very high temperature or in a very compact system. White dwarfs and neutron stars are the most compact and visible objects in our universe. Therefore, it is reasonable to expect that investigating the structure of such objects may uncover the fingerprints of quantum space-time \cite{40}.

The aim of this paper is to study the effects of the generalized uncertainty principle (\ref{1}) on the structure of an ideal white dwarf star. The constituents of an ideal white dwarf are assumed to be helium nuclei immersed in an electron gas. The source of the pressure holding up this ideal star from the gravitational collapse is assumed to be entirely due to the electron degeneracy pressure. But the most important contribution to its total mass comes from the helium nuclei. The electron gas in a white dwarf is so dense that it is assumed to be at the zero temperature. Therefore, we first derive the thermodynamical quantities of a Fermi gas obeying the generalized uncertainty principle (\ref{1}) at the zero temperature limit. The main property of this Fermi gas is that the density of particle number and the density of internal energy take finite values as the Fermi energy tends to infinity. It is a consequence of this fact that the number of accessible states of a particle obeying the generalized uncertainty principle (\ref{1}) in a finite volume is finite \cite{19}.
 Then we apply the equation of state of such a Fermi gas to the Newtonian hydrostatic equation which is describing the equilibrium configuration of a white dwarf. It is then proved that the radius of the star for all initial conditions has a finite value. We also show that the maximum mass of this ideal white dwarf, the Chandrasekhar limit, tends to infinity. It is the main result of this paper because for a conventional ideal white dwarf the Chandrasekhar limit has a finite value \cite{20}. In other words, a consequence of the existence of a minimal length scale is to remove the Chandrasekhar limit.

\section{Statistical mechanics of an ideal Fermi gas}

In this section we calculate the thermodynamic equations of an ideal Fermi gas which is in the ground state and obeying the generalized uncertainty relation (\ref{1}).

Although the commutation relations (\ref{a2}) and (\ref{a4}) imply the generalized uncertainty relation (\ref{1}) these relations alone do not completely specify the physical picture. To complete this picture we have to fix the behavior of these relations and quantities under Lorentz transformations. Assuming that the momentum $p_{\mu}$, the space-time coordinates $x^{\mu}$ and the wave-vector $k_{\mu}$ transform like standard 4-vectors under Lorentz transformations, the form of the commutation relations (\ref{a2}) and (\ref{a4}) does not remain invariant. It violates the principle of relativity. This case in which Lorentz symmetry is broken is strongly constrained by experimental data \cite{29}. However, these constraints are presently many order of magnitude away from the regime where expected that quantum gravity effects appear \cite{17}.

 Considering the validity of the principle of relativity, it is possible to modify the Lorentz transformation for the momentum and treat the wave vector and space-time coordinates as normal Lorentz 4-vectors in such a way that the commutation relations remain invariant (deformed special relativity)\cite{34,35}. In this approach, the linear sum of momenta is not a conserved quantity in all inertial frames. To solve this problem, one may define a new, non-linear, addition law that has the property of remaining invariant under nonlinear-modified Lorentz transformations. However, this non-linear addition law leads to the soccer-ball and the spectator problem \cite{36,37,38}.
Another possibility is that the momentum transforms as a standard Lorentz vector, while the wave vector and space-time coordinates transform under the nonlinear-modified Lorentz transformation in such a way that the commutation relations remain unchanged. In this case the mass-shell relation
\begin{equation}
\eta^{\mu\nu}p_{\mu}p_{\nu}-m^2=0,
\label{a6}
\end{equation}
is form invariant under the change of inertial frames of reference. This mass-shell relation can be treated as the Hamiltonian constraint of the theory. In this class of models, the speed of massless particles may depend on the momentum 4-vector. Dependence of the speed of light on momentum gives rise to a large macroscopic non-locality which causes serious conceptual problems and is incompatible with observation \cite{31,32,33}. To avoid this issue, one may abandon the notion of absolute locality and replace it by "the principle of relative locality" \cite{37,39}. However, it should be noted that this problem does not arise in all cases \cite{30}. For example, in two dimensional space-time, if one sets $p_{0}=f_{0}(k_{\nu})=g(k_{0})$ and $p_{1}=f_{1}(k_{\nu})=g(k_{1})$ in the massless limit, the speed of light becomes constant \cite{30}.

In the present paper we assume that the Energy-momentum vector $p_{\mu}$ as the physical momentum transforms like a normal Lorentz-vector under the change of inertial frames, whereas the wave vector and space-time coordinates transform under the nonlinear-modified Lorentz transformation in such a way that the commutation relations remain unchanged. In this case one can take the mass-shell relation (\ref{a6}) as the Hamiltonian constraint. For an isolated ideal gas as a closed and non-interacting composite system, the linear sum of all momenta of the individual particles is therefore a conserved quantity in all inertial frames. It can be then verified that the definitions of, and the relations between, various thermodynamical quantities all remain unchanged \cite{21}. In particular, the grand canonical partition function can be written in the form of the one-particle canonical partition function \cite{28}.
But, in conventional quantum statistical mechanics where $\lambda=0$,
due to the Heisenberg uncertainty principle, the phase space is
divided into cells of volume $h^3$ where $h$ is the Planck
constant. This means that the one-particle phase space measure is
given by $h^{-3}$ and therefore each integral on the phase space will be of
the form $\int d^3x d^3p h^{-3}(\Box)$. For a system obeying the GUP, the volume of phase space cells can be derived in several ways \cite{19,21,22,23,41}.
 All of them show that the phase
space should be divided into cells of volume $h^{3}(1+\lambda^2
p^2)^3$ and thus phase space integrals are of the form $\int d^3x d^3p
h^{-3}(1+\lambda^2 p ^2)^{-3}(\Box)$.
Therefore, the logarithm of the grand canonical partition function of a non-interacting system becomes
\begin{equation}
\ln \mathcal{Z}=\beta \mathcal{P}V=g\int\frac{d^3x d^3p
}{h^{3}(1+\lambda^2 p ^2)^{3}}\ln[1+e^{-\beta (\epsilon-\mu)}]  \label{2}
\end{equation}
where $V$ is the volume, $\beta$ is the inverse temperature, $\mathcal{P}$ is the pressure, $\mu$ is the chemical potential, $g$ is the degeneracy factor and $\epsilon$ is the eigen-value of the Hamiltonian of a noninteracting
Fermi particle and then the particle number of the system is
\begin{equation}
N=\frac{1}{\beta}\frac{\partial\ln \mathcal{Z}}{\partial\mu}|_{\beta,V}=g\int\frac{d^3x d^3p
}{h^{3}(1+\lambda^2 p ^2)^{3}}\frac{1}{1+e^{\beta (\epsilon-\mu)}}  \label{3}
\end{equation}
To investigate ideal white dwarfs one can typically assume that the temperature of the system is zero (i.e. the gas is in the ground state and completely degenerate) and the particles are ultra-relativistic  \cite{28}. Assuming $\epsilon=c p$, in the zero temperature limit equations (\ref{2}) and (\ref{3}) yield \cite{24}
\begin{eqnarray}
N&=&\frac{4\pi V g}{h^{3}}\int_{0} ^{p_{f}}\frac{ p^2dp
}{(1+\lambda^2 p ^2)^{3}}\nonumber\\
&=&\frac{\pi g V}{2h^3\lambda^3}(\frac{y^3-y}{(1+y^2)^2}+\tan^{-1}y)  \label{4}
\end{eqnarray}
and
\begin{eqnarray}
\mathcal{P}&=&\frac{4\pi g c}{h^{3}}\int_{0} ^{p_{f}}\frac{ p^2dp
}{(1+\lambda^2 p ^2)^{3}}(p_{f}-p)\nonumber\\
&=&\frac{\pi g}{2h^3\lambda^4}(\frac{-y^2}{1+y^2}+y \tan^{-1}y)  \label{5}
\end{eqnarray}
where $p_{f}=\mu/c$ is the momentum corresponding to the Fermi energy and $y=\lambda p_{f}$. Letting $\lambda$ goes to zero, the conventional equation of state for ideal Fermi gases is obtained.
If the value of the Fermi energy is taken to infinity, the particle number $N$ tends to the finite value $\frac{g \pi^2 V}{4 h^3 \lambda^3}$. It means that the total number of possible states of such a system is finite. This property is an integral part of a system obeying the generalized uncertainty relation  (\ref{1}) \cite{19}.

\section{Removing the Chandrasekhar limit}
If one considers a white dwarf as a Newtonian star with the conventional equation of state for ideal Fermi gases and takes the Fermi energy at the center of the star to infinity, the radius of star goes to zero and the mass of it tends to a finite value called the Chandrasekhar limit (or mass). Although the Chandrasekhar limit is considered as an upper bound on the  mass of an white dwarf, this limit is only a theoretical limit and before approaching it, actually when $p_{f}\sim5 m_{e} c$, electrons are captured by nuclei and turning protons into neutrons. The effect of this process is to increase the number of nucleons to electrons which will cause an instability and the star becomes a neutron star \cite{26}. Therefore, the maximum mass of a real white dwarf is less than the Chandrasekhar limit. Detailed calculations show that this mass is about $1.2$ times of the solar mass \cite{26}. Actually, at this energy level, i.e. $p_{f}\sim5 m_{e} c$, the effect of the minimal length is, in turn, negligible \cite{24,25}.

In this section we apply the modified equations of state (\ref{4}) and (\ref{5}) to the hydrostatic equation of a Newtonian star and show that when the central Fermi energy goes to infinity the radius and the mass of the star, as opposed to the conventional case, both tend to infinity. In other words these modified equation of states remove the Chandrasekhar limit.
The newtonian hydrostatic equation for an isotropic star is
\begin{equation}
\frac{d}{dr}(\frac{r^2}{\rho(r)}\frac{d\mathcal{P}(r)}{dr})+4\pi G r^2\rho(r)=0,  \label{6}
\end{equation}
where $r$ is the distance from the center of star, $\rho(r)$ and $\mathcal{P}(r)$ are the mass density and the pressure at $r$ respectively. Since in white dwarfs the mass of electrons is very small compared to the mass of nucleons, the mass density can be written as
\begin{equation}
\rho(r)=\mu m_{N} n(r), \label{7}
\end{equation}
where $n$ is the number density of electrons, $\mu$ is the number of nucleons per electron and $m_{N}$ is the mass of a nucleon. The temperature of a white dwarf is sufficiently low such that the electrons will be frozen at the lowest energy levels. Using (\ref{4}), (\ref{5}) and (\ref{7}), the differential equation (\ref{6}) yields
\begin{equation}
k\frac{d}{dr}(r^2\frac{dy(r)}{dr})+4\pi G r^2 \rho(y(r))=0, \label{8}
\end{equation}
where $k=c/\mu m_{N}\lambda$. To solve this differential equation we need two initial conditions. The physical reasonable initial conditions for an isotropic star are
\begin{equation}
y(r=0)=y_{c} \hspace{0.5 cm},\hspace{0.5 cm} \frac{d y}{dr}(r=0)=0, \label{9}
\end{equation}
where $y_{c}$ is the Fermi energy of the center of the star.
If we set
\begin{eqnarray}
\Theta(r)&\equiv& y(r)/y_{c},\nonumber\\
\rho_{m}&\equiv& \lim_{y\rightarrow\infty}\rho(y)=\frac{\pi^2 g \mu m_{N}}{4\lambda^3h^3},\nonumber\\
\rho_{e}&\equiv& \rho(y)/\rho_{m}=\frac{2}{\pi}(\frac{y^3-y}{(1+y^2)^2}+\tan^{-1}(y)),\nonumber\\
\xi &\equiv& \sqrt{\frac{4\pi G \rho_{m}}{k y_{c}}}\hspace{.2 cm}r \geq 0  \label{10}
\end{eqnarray}
and substitute them in equation (\ref{8}), we get
\begin{equation}
\frac{1}{\xi^2}\frac{d}{d\xi}(\xi^2\frac{d\Theta(\xi)}{d\xi})+\rho_{e}(y_{c}\Theta(\xi))=0. \label{11}
\end{equation}
The initial conditions, then, become
\begin{equation}
\Theta(0)=1 \hspace{0.5 cm},\hspace{0.5 cm} \frac{d\Theta}{d\xi}(0)=0. \label{12}
\end{equation}
This equation is a generalized form of the Lane-Emden equation and, as expected, reduces to the Lane-Emden equation at the zero $\lambda$ limit. It has been proved that this initial value problem has a unique solution \cite{27}. Equation (\ref{11}) as a differential equation describing the configuration of a star is acceptable if we prove that the radius of star determined by this equation always has a finite value.

 In fact, It is possible to prove that for any $y_{c}>0$, the solution satisfying equation (\ref{11}) and the initial conditions (\ref{12}) has at least a finite positive root. We will refer to the smallest one as $\xi_{R}$. Therefore, the radius of star always has a finite value
\begin{equation}
R=\sqrt{\frac{k y_{c}}{4\pi G\rho_{m}}}\hspace{.2 cm}\xi_{R}. \label{13}
\end{equation}
 The total stellar mass is then determined by
\begin{eqnarray}
M&=&4\pi\int_{0}^{R}r^2\rho(r)dr\nonumber\\
&=&(4\pi\rho_{m})^{-\frac{1}{2}}(\frac{k y_{c}}{G})^{\frac{3}{2}}\int_{0}^{\xi_{R}}\xi^{2}\rho_{e}(y_{c}\Theta(\xi))d\xi\nonumber\\
&=&-(4\pi\rho_{m})^{-\frac{1}{2}}(\frac{k y_{c}}{G})^{\frac{3}{2}}\hspace{.2 cm}\xi_{R}^{2}\frac{d\Theta}{d\xi}(\xi_{R}), \label{14}
\end{eqnarray}
 where the last equality is obtained by using the differential equation (\ref{11}).

 Before proving the above claim we need to prove two very simple lemmas.

 \textbf{Lemma 1}. Let $\Theta(\xi)$ be the solution of equation (\ref{11}). Then $\Theta(\xi)$ is a decreasing function on the interval $[0,\xi_{R})$ for any $y_{c}>0$ and consequently is always less than or equal to unity on the interval $[0,\xi_{R})$.

 It should be noted that we don't assume that $\xi_{R}$ has a finite value at this stage and it is possible for $\xi_{R}$ to be infinity.
 Since $\Theta(0)=1>0$ and $\xi_{R}$ is the smallest positive root of $\Theta(\xi)$ the function $\Theta(\xi)$ must be positive every where on $[0,\xi_{R})$ by continuity. It implies that $\rho_{e}(y_{c}\Theta(\xi))\geq 0$ for all $y_{c}>0$ and $\xi_{R}\geq\xi\geq 0$. By using the differential equation (\ref{11}) one can deduce that $(\xi^2\frac{d\Theta(\xi)}{d\xi})$ is a decreasing function and since $(\xi^2\frac{d\Theta(\xi)}{d\xi})\mid_{0}=0$ the function $\frac{d\Theta(\xi)}{d\xi}$ must be a negative function; hence $\Theta(\xi)$ is a decreasing function.

\textbf{Lemma 2}. If $0\leq\Theta\leq1$, then for every $y_{c}>0$ there exists a positive real number $a$ depending only on $y_{c}$ such that
\begin{equation}
\rho_{e}(y_{c}\Theta)\geq a (y_{c}\Theta)^3. \label{15}
\end{equation}
For example, $a=\frac{1}{3\pi(1+y^2)^3}$ satisfies the above inequality.

Now, we are ready to establish the proof of the following theorem.

 \textbf{Theorem 1}. For any $y_{c}>0$, the solution of the initial value problem (\ref{11}) and (\ref{12}) has a finite positive root.

 Setting $u(\xi)\equiv\xi\Theta(\xi)$ and substituting it to equation (\ref{11}), we have
 \begin{equation}
\frac{d^2 u(\xi)}{d\xi^{2}}+\xi\rho_{e}(y_{c}\frac{u(\xi)}{\xi})=0. \label{16}
\end{equation}
It is a differential equation corresponding to a nonlinear oscillator with initial condition
 \begin{equation}
u(0)=0 \hspace{0.5 cm},\hspace{0.5 cm} \frac{d u}{d\xi}(0)=1. \label{17}
\end{equation}
It is enough to prove that the solution of (\ref{16}) has a finite positive root for every $y_{c}>0$.

Suppose by contradiction that there exists a positive $y_{c}$ such that $u(\xi)>0$ for all $\xi>0$. It implies that $\rho_{e}(y_{c}\frac{u(\xi)}{\xi})>0$. From the differential equation (\ref{16}) it follows that $\frac{d^2u(\xi)}{d\xi^{2}}<0$ for all $\xi>0$ and therefore $\frac{d u(\xi)}{d\xi}$ is a monotonically decreasing function on $(0,\infty)$. But, $\frac{d u(\xi)}{d\xi}$ will not tend to a negative limit as $\xi\rightarrow\infty$ because this would imply that $u(\xi)$ is ultimately negative. Then we have $0\leq\lim_{\xi\rightarrow\infty}\frac{d u(\xi)}{d\xi}\leq1$.

Integrating (\ref{16}) over $(0,\xi)$, we obtain
\begin{equation}
 \frac{d u}{d\xi}(\xi)-\frac{d u}{d\xi}(0)+\int_{0}^{\xi}\xi'\rho_{e}(y_{c}\frac{u(\xi')}{\xi'})d\xi'=0.  \label{18}
\end{equation}
The integral on the left of (\ref{18}) converges as $\xi\rightarrow\infty$ because $\frac{d u}{d\xi}(\xi)$ tends to a finite value at this limit. Therefore, one can integrate (\ref{16}) over $(\xi,\infty)$, getting now
\begin{equation}
 \frac{d u}{d\xi}(\infty)-\frac{d u}{d\xi}(\xi)+\int_{\xi}^{\infty}\xi'\rho_{e}(y_{c}\frac{u(\xi')}{\xi'})d\xi'=0,  \label{19}
\end{equation}
Since $\frac{d u}{d\xi}(\infty)\geq 0$, equation (\ref{19}) implies
\begin{equation}
 \frac{d u}{d\xi}(\xi)\geq\int_{\xi}^{\infty}\xi'\rho_{e}(y_{c}\frac{u(\xi')}{\xi'})d\xi'.  \label{20}
\end{equation}
Integrating the above equation, we get
\begin{eqnarray}
 u(\xi)-u(0)&\geq&\int_{0}^{\xi}d t\int_{t}^{\infty}\xi'\rho_{e}(y_{c}\frac{u(\xi')}{\xi'})d\xi'\nonumber\\
 &\geq&\int_{0}^{\xi}\xi'^{2}\rho_{e}(y_{c}\frac{u(\xi')}{\xi'})d\xi'\nonumber\\
 &+&\xi\int_{\xi}^{\infty}\xi'\rho_{e}(y_{c}\frac{u(\xi')}{\xi'})d\xi',  \label{21}
\end{eqnarray}
which gives
\begin{equation}
 u(\xi)\geq\int_{0}^{\xi}\xi'^{2}\rho_{e}(y_{c}\frac{u(\xi')}{\xi'})d\xi'.  \label{22}
\end{equation}
According to Lemma 1, we have $0\leq\frac{ u(\xi)}{\xi}=\Theta(\xi)\leq1$ for all $\xi\geq0$. Applying Lemma 2, the inequality (\ref{22}) yields
\begin{equation}
 u(\xi)\geq(a y^3)\int_{0}^{\xi}\frac{u(\xi')^3}{\xi'}d\xi',  \label{23}
\end{equation}
which implies
\begin{equation}
 \frac{u^3(\xi)/\xi}{(\int_{0}^{\xi}\frac{u^3(\xi')}{\xi'})d\xi')^3}\geq\frac{(a y^3)^3}{\xi}  \label{24}
\end{equation}
for all $\xi\geq0$.
Assuming $0<\xi_{1}<\xi_{2}$ and integrating (\ref{24}) over the interval $(\xi_{1}, \xi_{2})$, we obtain
\begin{eqnarray}
 [(\int_{0}^{\xi_{1}}\frac{u^3(\xi')}{\xi'})d\xi')^{-2}
 &-&(\int_{0}^{\xi_{2}}\frac{u^3(\xi')}{\xi'})d\xi')^{-2}]\nonumber\\
 &\geq&2(a y^3)^3\int_{\xi_{1}}^{\xi_{2}}\xi^{-1}d\xi.  \label{25}
\end{eqnarray}
Taking $\xi_{2}$ to infinity, the left hand side takes a finite value but the right hand side goes to infinity which shows a contradiction and the proof is complete.

To investigate the Chandrasekhar limit we should take the central Fermi momentum $y_{c}$ to infinity and calculate the total stellar mass. In the conventional case the mass approaches a finite value but, in contrast to this well known case, it is possible to show that for a white dwarf star governed by the hydrostatic equilibrium equation (\ref{8}) the Chandrasekhar limit tends to infinity. On the other words, the effect of quantum gravity removes the Chandrasekhar limit.

 Taking the central Fermi momentum $y_{c}$ to infinity, $\rho_{e}(y_{c}\Theta)$ tends to unity for all $\Theta>0$. So at this limit the differential equation (\ref{11}) converges to
 \begin{equation}
\frac{1}{\xi^2}\frac{d}{d\xi}(\xi^2\frac{d\Theta(\xi)}{d\xi})+1=0 \label{26}
 \end{equation}
with the solution
 \begin{equation}
 \Theta_{\infty}(\xi)=-\frac{\xi^{2}}{6}+1 \label{27}
 \end{equation}
and $\xi_{R}$ then goes to $\sqrt{6}$. Using equations (\ref{13}) and (\ref{14}), one can infer that the radius $R$ and the total mass $M$ both tend to infinity at the infinite $y_{c}$ limit.
Although, it is not difficult to prove that the solution of (\ref{11}), i.e. $\Theta_{y_{c}}(\xi)$, converges to $\Theta_{\infty}(\xi)$ as $y_{c}\rightarrow\infty$, we do not actually require to do this for determining the chandrasekhar limit. It is only enough to prove the following Lemma.

 \textbf{Lemma 3}. Let $\Theta_{y_{c}}(\xi)$ and $\Theta_{\infty}(\xi)$ be the solutions of equations (\ref{11}) and (\ref{26}) with the initial conditions (\ref{12}) respectively. Then, for any $y_{c}>0$, $\Theta_{y_{c}}(\xi)\geq\Theta_{\infty}(\xi)$ if $0\leq\xi\leq\xi_{R}$.

 Subtracting equation (\ref{26}) from equation (\ref{11}), we get
\begin{equation}
\frac{1}{\xi^2}\frac{d}{d\xi}(\xi^2\frac{d}{d\xi}(\Theta_{y_{c}}(\xi)-\Theta_{\infty}(\xi)))
=1-\rho_{e}(y_{c}\Theta_{y_{c}}(\xi)). \label{28}
 \end{equation}
Since $\rho_{e}(y_{c}\Theta_{y_{c}}(\xi))\leq1$, the function $\xi^2\frac{d}{d\xi}(\Theta_{y_{c}}(\xi)-\Theta_{\infty}(\xi))$ is an increasing function and according to the initial conditions (\ref{12}) we have $\frac{d}{d\xi}(\Theta_{y_{c}}(\xi)-\Theta_{\infty}(\xi))\geq0$ which implies that $(\Theta_{y_{c}}(\xi)-\Theta_{\infty}(\xi))\geq0$.

It also follows from this lemma that $\xi_{R}(y_{c})\geq\xi_{R}(\infty)=\sqrt{6}$.
By virtue of this Lemma and since $\rho_{e}$ is an increasing function, we find that
\begin{eqnarray}
M(y_{c})&=&4\pi\int_{0}^{R}r^2\rho(r)dr\nonumber\\
&=&(4\pi\rho_{m})^{-\frac{1}{2}}(\frac{k y_{c}}{G})^{\frac{3}{2}}\int_{0}^{\xi_{R}(y_{c})}\xi^{2}\rho_{e}(y_{c}\Theta_{y_{c}}(\xi))d\xi\nonumber\\
&\geq&(4\pi\rho_{m})^{-\frac{1}{2}}(\frac{k y_{c}}{G})^{\frac{3}{2}}\int_{0}^{\xi_{R}(\infty)}\xi^{2}\rho_{e}(y_{c}\Theta_{\infty}(\xi))d\xi\nonumber\\ \label{29}
\end{eqnarray}
To find the limiting mass, the Chandrasekhar mass, we need to increase the central Fermi momentum $y_{c}$ without bound. At this limit $\rho_{e}(y_{c}\Theta_{\infty}(\xi))$ tends to unity and then the last integral in the above inequality converges to $2\sqrt{6}$ ; hence we have
\begin{equation}\label{30}
\lim_{y_{c}\rightarrow\infty}M(y_{c})\geq(4\pi\rho_{m})^{-\frac{1}{2}}\lim_{y_{c}\rightarrow\infty}(\frac{k y_{c}}{G})^{\frac{3}{2}}=\infty.
\end{equation}
Therefore, the maximum mass of an ideal white dwarf can increase without bound. Since the Chandrasekhar mass has a finite value in the conventional case it is reasonable to infer that this result is a pure effect of a quantum gravity which induces a minimal length scale in quantum mechanics.
Here an important question arises, whether this result is sensitive to the model that we choose to implement a minimal length scale in quantum mechanics or not. The main property of the system, which plays the crucial role to achieve this result, is that both the density function $\rho_{e}$ and the particle density $n$ take finite values as the Fermi momentum $y$ tends to infinity and the details of these functions are not important. It means that the number of accessible states of a particle in a finite volume is finite \cite{19}. Therefore, it seems that the result is independent of the model.

 Here, it is of importance to address the stability of the star. To show the stability of the configuration for all $y_{c}$, it is enough to prove that the total mass $M$ as a function of $y_{c}$ is always an increasing function. Although, unfortunately, we haven't found any exact proof yet, numerical calculations show that it is probably the case. It is still an open problem for us.

Lastly, we should add a remark concerning the correction to the mass of white dwarfs which has been derived in \cite{24}. To acquire the correction, it is assumed that
\begin{equation}\label{31}
\mathcal{P}(R)=\frac{\alpha}{4\pi}\frac{G M^2}{R^4},
\end{equation}
where $\mathcal{P}$ is the pressure and $\alpha$ is a number whose exact value depends on the distribution of matter inside stars. Then, an approximate relation between the pressure and the number density has been employed to obtain a corrective term for the total mass. The correction is legitimate for a typical white dwarf because, as mentioned earlier, the maximum mass of a real white dwarf is around $1.2$ times of the solar mass and at this energy level, the above approximations are acceptable but for an ideal white dwarf whose mass is close to the Chandrasekhar limit, equation (\ref{31}) is no longer valid. A simple calculation shows that equation (\ref{31}) holds only for Newtonian polytropes whose equations of state take the form $\mathcal{P}\sim\rho^{\gamma}$ (e.g. see the page $311$ of \cite{26}). Therefore, as shown in the present paper, the effect of a minimal length on the Chandrasekhar limit cannot be simply reducible to a corrective term.
\section{Summary and Conclusion}
We investigated the quantum gravity influences on the equilibrium configuration of a white dwarf star. We assumed a quantum gravity which induces a minimal length scale in quantum mechanics. This effect can be implemented as a deformation in the Heisenberg algebra between the position operator and the momentum operator. Since a white dwarf is a very compact object we typically assumed that the ideal Fermi gas constituting the star is in the ground state and completely degenerate. Then the thermodynamic equations of such a system were obtained. The main feature of this system is that the density of particle number and the density of internal energy always take finite values, even if the Fermi energy tends to infinity. It is a consequence of this fact that the number of accessible states of a particle obeying the generalized uncertainty principle (\ref{1}) in a finite volume is finite \cite{19}.

Applying the equation of state of the Fermi gas to the Newtonian hydrostatic differential equation of an isotropic star, we obtained a generalized form of the Lane-Emden equation. We then showed that the radius of the star whose configuration is described by this generalized Lane-Emden equation always has a finite value. Finally, we focused on the Chandrasekhar mass, the maximum mass of an ideal white dwarf star, and proved that the value of the maximum mass of the star is infinite. This is in direct contradiction to the conventional case where the Chandrasekhar mass has a finite value. An important thing that should be noted here is that for a real white dwarf the Chandrasekhar limit is not actually attainable because of the electron capture effect.
Before reaching the Chandrasekhar limit, electrons are captured by nuclei and turning protons into neutrons. The effect of this process is to increase the number of nucleons to electrons and then the electron degeneracy pressure can no longer prevent the star from collapsing under the gravitational force. Therefore, the core of the star collapses to become a neutron star \cite{26}. Since the gravitational field of a neutron star is very strong it is reasonable to use the general theory of relativity and the Tolman-Oppenheimer-Volkoff hydrostatic equation rather than the Newtonian hydrostatic equation. The question then arises whether or not the Oppenheimer-Volkoff limit, the maximum mass of a stable neutron star, is avoidable in the presence of a minimal length scale. We will deal with this problem in future works.

\section{Acknowledgments}
The author is very thankful to Dr. M. Pourbarat for helpful and enlightening discussions.

\end{document}